\documentstyle[11pt,emulateapj,newapa]{article}  
\catcode`\@=11 
\def\lsim{\mathrel{\mathpalette\@versim<}}
\def\gsim{\mathrel{\mathpalette\@versim>}}
\def\be{\begin{equation}}
\def\ee{\end{equation}}

\def\K{\rm K}
\def\keV{\rm \, keV} 
\def\kpc{\rm \, kpc} 
\def\rtr{r_{\rm tr}}
\def\Rtr{R_{\rm tr}}
\def\Rg{R_{\rm g}}
\def\mdot{\dot{m}}
\def\Mdot{\dot{M}}
\def\mdotcr{\dot{m}_{\rm crit}}
\def\Mdotedd{\dot M_{\rm Edd}}
\def\msun{M_{\odot}}
\def\a0{A0620$-$00}
\def\nm91{GRS~1124$-$68}
\def\@versim#1#2{\vcenter{\offinterlineskip
        \ialign{$\m@th#1\hfil##\hfil$\crcr#2\crcr\sim\crcr } }}
\catcode`\@=12 

\submitted{To Appear in The Astrophysical Journal}
\righthead{Esin et al.}
\lefthead{Optical Lightcurves of \nm91 and \a0 in Outburst}

\eqsecnum

\begin{document}
\date{}  
\title{Optical Lightcurves of the Black Hole Binaries \nm91 and \a0 in 
Outburst: the Importance of Irradiation}
\author{Ann A. Esin\altaffilmark{1}} 
\altaffiltext{1}{Chandra Fellow}
\affil{Caltech 130-33, Pasadena, CA 91125; aidle@tapir.caltech.edu}
\affil{Institute for Theoretical Physics, University of California, 
Santa Barbara, CA 93106}
\author{Erik Kuulkers}
\affil{Space Research Organization Netherlands, Sorbonnelaan 2, 3584 CA
    Utrecht, The Netherlands; E.Kuulkers@sron.nl}
\affil{Astronomical Institute, Utrecht University, P.O. Box 80000, 
    3507 TA Utrecht, The Netherlands}
\and
\author{Jeffrey E. McClintock and Ramesh Narayan}
\affil{Harvard-Smithsonian Center for Astrophysics, 60 Garden Street,
Cambridge, MA 02138; \\ jmcclintock@cfa.harvard.edu,
rnarayan@cfa.harvard.edu}

\begin{abstract} 
We test whether the model proposed by Esin~et~al. to explain X-ray
observations of the black hole soft X-ray transient 
\nm91 in outburst can also explain the optical
lightcurves of a similar object \a0.  We show that
to reproduce the observed X-ray to optical flux ratio in \a0, we need
to assume X-ray irradiation of the outer disk that is significantly in
excess of what is expected for a standard planar disk.  With enhanced
irradiation, the Esin~et~al. model can reproduce the optical evolution
of \a0 in outburst very well.  Though we find that optical
observations of \nm91 also imply enhanced X-ray heating of the outer
disk, the irradiation appears to be a factor of $\sim 3$ weaker in this
system than in \a0.  This is surprising, since \nm91 has a 
larger disk.  We speculate that enhanced irradiation may be due to
disk warping, and that the degree of warping differs between the
two binaries.
\end{abstract} 


\section{Introduction}
\label{intro}

A number of known X-ray sources exhibit dramatic outbursts in which
their bolometric luminosity increases by many orders of magnitude.  A
fraction of these systems, termed soft X-ray transients (SXTs), is
believed to contain binaries in which a more massive compact star is
accreting mass from a less massive late-type companion (see
\fcitep{tas96} and \fcitep{vpm95} for the most recent comprehensive
reviews, though several new SXTs were discovered in the last few
years).  This subset of X-ray transients is especially interesting
because the low mass of their secondary companions has in several
cases allowed us to measure the mass functions of the compact primary
to be above $3\msun$, thus providing the first (and so far only)
unambiguous evidence for the existence of stellar mass black holes (see
\fcitep{mcc98} and \fcitep{cha99} for the most up-to-date list of
black hole SXTs).

The behavior of black hole SXTs (BHSXT) during their outbursts shows a
number of common features.  A ``typical'' lightcurve consists of a
fast rise ($\tau \sim$ few days) followed by a longer ($\tau \sim$
month) exponential decline back to the system's quiescent luminosity,
sometimes with one or more secondary maxima, though not all BHSXTs
follow this standard prescription (see \fcitep{csl97} for an excellent
inventory of observed lightcurves).  The X-ray spectral evolution of
these systems during outbursts is less well studied.  There are many
indications that the majority of sources start out fairly soft near
the peak (hence their name), i.e. with the total emission dominated by
a component with $T \sim$ few keV, and soften further during the
decline, before becoming suddenly very hard (the total energy output
is dominated by $\sim 100\keV$ photons) several months after the peak
(e.g. \scitep{tal95}; \scitep{tas96}).  However, there are several
exceptions; e.g. GS~2023$+$338 and GRO~J0422$+$32 remained hard
throughout their outburst, though the latter displayed a very typical
exponential lightcurve.  Very little is known about the X-ray spectra
of BHSXTs in quiescence, due mainly to their faintness; only V404 Cyg
has a well determined spectrum (see \fcitep{nbm97}).

It is now generally accepted that the outbursts in BHSXTs are caused
by a thermal instability in the accretion disk, the same mechanism
that is believed to cause outbursts in dwarf novae (see
e.g. \fcitep{can93}; \fcitep{las96}; \fcitep{can98}).  This model
relies on the fact that in an optically thick accretion disk with a
central temperature $T_c \sim 10^4\,\K$ (the ionization temperature of
hydrogen) the opacity is a very strong function of $T_c$.  Both cold
neutral disks and hot fully-ionized disks are stable, but when the
central disk temperature is close to the critical value, thermal
instability forces the disk to undergo a limit cycle.  In quiescence
the disk is cold and its mass accretion rate is considerably below the
mass transfer rate from the companion.  In this way the disk
accumulates mass until increasing density and temperature force it
into a hot ionized state, characterized by a very high (often on the
order of $\Mdotedd$) mass accretion rate, thereby starting an
outburst.  Of course this high mass accretion rate can only be
sustained until the surface density in the disk is depleted
sufficiently to force a transition back into the cold neutral state,
after which the system settles back into quiescence.

A detailed numerical model of such an outburst in a BHSXT is currently
under construction (see e.g. \fcitep{dlh99}; \fcitep{hmd98}).  It is
already clear however, that illumination of the outer disk by X-rays
produced near the black hole plays a very important role
(e.g. \fcitep{pam94}; \fcitep{par96}; \fcitep{las99}).  \fcitet{kir98}
in a recent paper plausibly argued that such irradiation would
naturally produce a prototypical exponential lightcurve.  These
authors also attribute the departures from the standard lightcurve
shape to the size of the disk, claiming that in larger systems
irradiation is unable to heat up the outer disk to temperatures above
$10^4\,\K$.  In such a case a cooling wave setting off from the outer
disk edge would shut off the mass flow early on, producing a short
outburst with a linear decline.  This suggestion seems to be supported
by observations, as shown by \fcitet{sck98}, as well as by detailed
theoretical modeling (\scitep{dlh99}).

Though the disk instability model outlined above offers a plausible
explanation of the general lightcurve shapes seen in outbursts of
BHSXTs, the origins of the spectral evolution of these systems are
still very much debated in the literature.  In an attempt to unify
different ``spectral states'' seen during the decline from the
outburst peak to quiescence, \facite{emn97} (1997, hereafter EMN) have
proposed a model for the prototypical BHSXT \nm91 (also Nova Muscae
1991).  In their scenario, luminous soft spectra (with characteristic
photon index $\Gamma \lsim 2.5$) seen near the peak of the outburst,
when the system is in the so called Very High and High states
(hereafter VHS and HS respectively), are modeled by a cool optically
thick accretion disk (\fcitep{shs73}) topped with a hot optically thin
corona (e.g. \fcitep{ham91}).  In the less luminous Low state
(hereafter LS), when the observed spectra are very hard (with $\Gamma
\sim 2.0-1.5$), as well as in the Quiescent state, the accretion flow
assumes a different configuration: a hot and optically thin inner
flow (a sort of diskless corona) is surrounded by an optically thick
outer disk with an inner edge anywhere between several tens and tens
of thousands Schwarzschild radii from the accreting black hole.

\acite{emn97} model the hot inner part as an optically thin
advection-dominated accretion flow (ADAF) discovered by
\fcitet{ich77}, which has been extensively discussed in the literature
in the last six years (see \fcitep{nmq98} and \fcitep{kfm98} for
latest reviews).  An ADAF is a hot and stable accretion flow solution
alternative to the standard \fcitet{shs73} thin disk.  It is
characterized by electron temperatures above $10^9\,\K$ and optical
depth $\lsim 1$, which are ideal conditions for producing the hard
spectra seen in the low state through thermal Comptonization of the
seed synchrotron and thin disk photons.  Since an ADAF solution only
exists for mass accretion rates below some critical value, $\Mdot_{\rm
crit}$, it suggests a natural explanation for the transition from the
hot flow $+$ outer thin disk configuration at low $\Mdot$ to thin disk
$+$ corona at $\Mdot > \Mdot_{\rm crit}$.

A model involving an advection-dominated flow at accretion rates well
below $\Mdot_{\rm crit}$ surrounded by an outer thin disk had already
been shown to explain well the puzzling properties of quiescent BHSXTs
(e.g. \fcitep{nmy96}; \scitep{nbm97}; \fcitep{hlm97}).  In their paper,
\acite{emn97} demonstrated that the scenario outlined above works 
for systems in outburst; it reproduces
very well both soft and hard X-ray lightcurves of \nm91 as well as the
details of its spectral evolution.  However, though \nm91 had
excellent X-ray coverage, the optical (e.g. \fcitep{bai92};
\fcitep{khm96}; \fcitep{dmb98}) and UV (\fcitep{sgr93}) observations,
which are the only diagnostic of the outer disk properties, were
limited to the first 150 days of the outburst.  At later times EMN
could obtain useful constraints only on the innermost regions of the
accretion flow, the properties of which are well constrained by the
observed X-ray emission.  Particularly unfortunate is the fact that
the important High--Low state transition occurred during a period
with almost no optical coverage.  According to the EMN model, during this
transition, the source of the X-ray emission shifted from the flat
disk to a quasi-spherical corona, causing a considerable change in the
angular distribution of radiation incident onto the outer disk.  This
effect could potentially be seen in the optical light curve.

Though we cannot remedy the lack of optical coverage for \nm91, in
this paper we attempt to do the next best thing, i.e. we attempt to
complement the work of \acite{emn97} by using the extensive
multi-wavelength photometry available for the outburst of another
BHSXT, \a0, to infer the properties of the outer disk in these two
systems.  The quality of the X-ray data available for \a0 (though
superior for its time; see \scitep{kuu98}) renders it unusable for
detailed modeling similar to that done for \nm91.  However, in
sections \S\ref{binary} and \S\ref{xrays} we will demonstrate that the
two systems are sufficiently alike in their binary parameters and
outburst properties to justify a direct use of the model proposed for
\nm91, with only minor parameter changes.

In \S\ref{opt} and \S\ref{optnm} we describe the modeling of the
optical data and present our results, which are then further discussed
in \S\ref{disc}.  We conclude with a summary in \S\ref{conc}.

\section{\a0 and \nm91 as Binary Systems}
\label{binary}

The remarkable similarity between \nm91 and \a0 in their orbital
parameters (\fcitep{bai92}; \fcitep{rmb92}), appearance of the
secondary (\scitep{rmb92}) and of the accretion disk in quiescence
(\scitep{obr94}), 
has been well documented in the literature.  To emphasize how closely
these two systems resemble each other, we show in Table
\ref{tabbin} a summary of their latest binary parameters, gathered
from the literature.

\placetable{tabbin}

The properties of BHSXT outburst lightcurves depend largely on the
mass of the primary, which determines the overall energy scale for the
system, mass transfer rate from the companion, and the size of the
accretion disk.  The latter is very important since it probably
dictates the value of the mass accretion rate at the peak of the
outburst as well as the overall lightcurve shape
(e.g. \scitep{kir98}).  It is clear from Table \ref{tabbin}, that the
data are consistent with the black holes in \nm91 and \a0 having similar
masses.  Since their orbital periods are short, this argues that the
mass transfer rate cannot be very different in the two systems
(e.g. \fcitep{kkb96}).  Finally, the size of the accretion disk,
$R_{\rm out}$, can be estimated from other orbital parameters,
assuming that it fills a fixed fraction of the primary Roche Lobe (we
take it to be $80\%$).  We compute $R_{\rm out}$ using the
\cite{pac71} formula:
\begin{eqnarray}
\label{rout}
R_{\rm out} &=& 2.82\times 10^{10} P_{\rm orb}^{2/3} 
[m (1+q)]^{1/3} \\ \nonumber
&\times& (0.38-0.2 \log{q})\ {\rm cm},
\end{eqnarray}
where $P_{\rm orb}$ is the binary orbital period measured in hours, 
$q=M_c/M$ is the
mass ratio, and $m=M/\msun$ is the mass of the black hole in solar
units.  In calculating accretion disk size for \a0 we adopted $m=4.5$
(see \S\ref{xrays}) and $q=0.067$, while for \nm91 we took $m=6$ and
$q=0.133$.  The resulting values of $R_{\rm out}$ for both objects
with the corresponding uncertainties are shown in Table \ref{tabbin}.
Since the orbital period of \a0 is shorter than that of \nm91, the
accretion disk of the former is somewhat smaller in physical units,
though in Schwarzschild units the two disks have roughly the same sizes.

\section{X-Ray Outbursts of \nm91 \& \a0}
\label{xrays}

Given that these two binaries resemble each other so closely, it is
not surprising that their properties in outburst are similar as well,
as we demonstrate below.  In Figure \ref{counts} (upper panels) we
have plotted the observed soft X-ray lightcurves of \a0 and \nm91.  Both
systems show a fast rise and a characteristic exponential decline
after the peak, interrupted by a small secondary and a somewhat larger
tertiary maximum.  (Though it is not clear from this data that \a0
reaches a local maximum near day 200, other instruments showed that
the X-ray flux indeed declined shortly after that point, see
\scitep{kuu98}.)  The rates of the exponential decline after the
primary maximum in the two objects are very close; for \a0 we have
$\tau \sim 32\,{\rm days}$, and for \nm91, $\tau \sim 35\,{\rm days}$.

It is considerably more difficult to compare spectral evolutions of
the two BHSXTs, since all published spectra of \a0 were taken during
the first three months after the onset of the outburst.  These
observations all show fairly soft spectra with characteristic peak
temperatures around a few keV (e.g. \fcitep{ces76}; \fcitep{ccd76};
\fcitep{dea76}; \fcitep{lok78}) confirming that this system indeed
looked like \nm91 in the VHS and HS.  After
about $\sim 100$ days, we have to rely on the broadband data, so in
Figure \ref{counts} (lower panels) we have plotted ratios of count rates in
the hard and soft X-ray bands for both objects.  The figure shows a
nearly identical slow decrease of the hardness ratios for both
systems, followed by an abrupt hardening just prior to the rise to the
tertiary maximum.  The latter feature suggests that a High--Low state
transition is taking place.

\centerline{\null}
\includegraphics{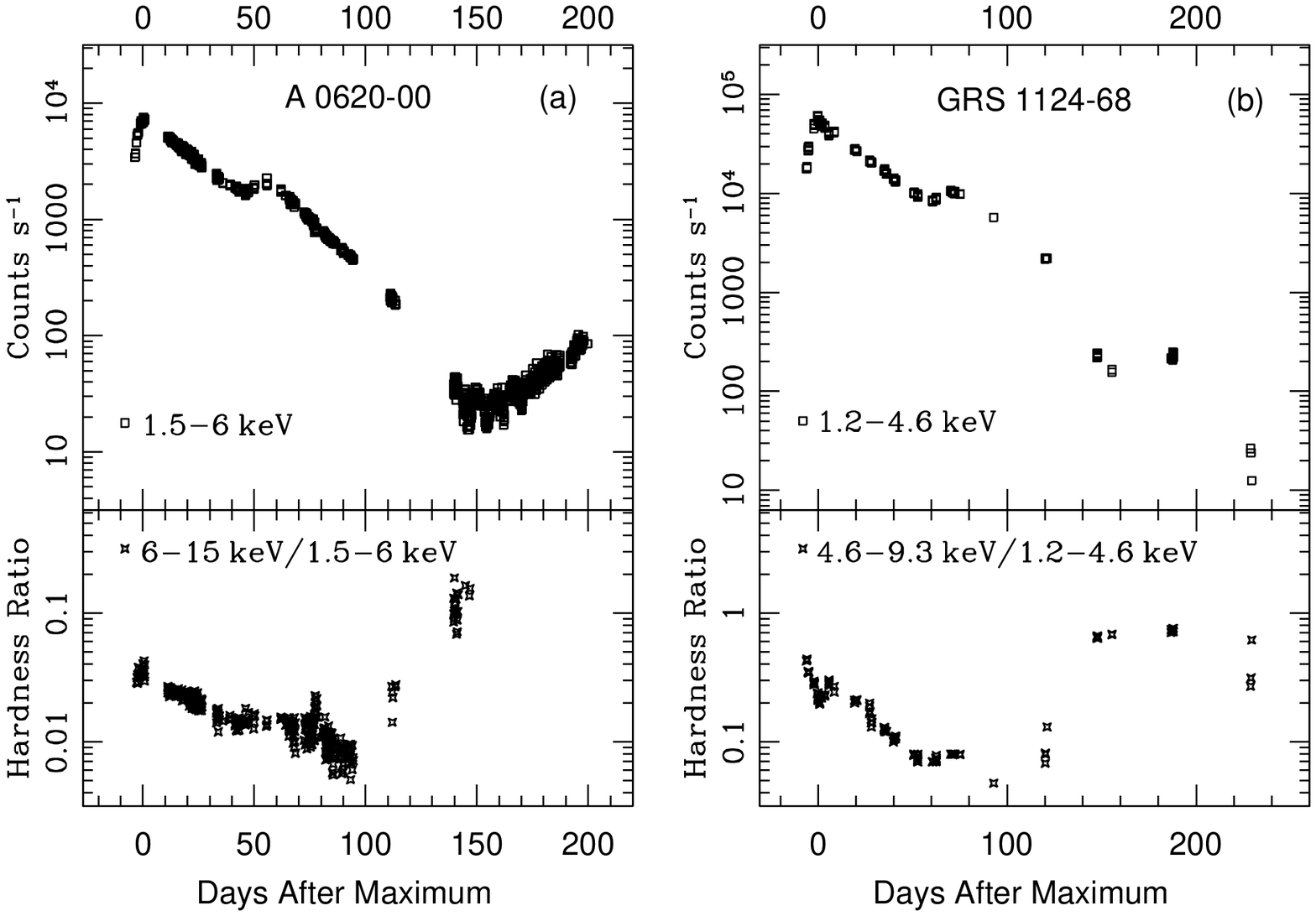} 
\vskip 2.5in 
\figcaption{\label{counts}(a) Soft X-ray lightcurve and hardness
ratio of \a0 obtained with {\it SAS-3} $CSL\,A$ ($1.5-6\keV$) and
$CSL\,B$ ($6-15\keV$) detectors (courtesy K. Plaks, J. Woo and
G. Clark; adopted from Kuulkers 1998). (b) Soft X-ray lightcurve and 
hardness ratio of \nm91 obtained with {\it Ginga LAC} detectors 
(courtesy of K. Ebisawa).  Note that error bars are not plotted in 
either panel.}
\vskip 0.5cm

So far in our comparison of the X-ray lightcurves and hardness ratios
we used raw count rates, since this is the information most readily
available for \a0.  Though in general a questionable procedure, we
believe it is justified here because {\it Ginga LAC} and {\it SAS-3}
contained detectors with very similar responses in the energy
bands of interest (\fcitep{buf77}; \fcitep{tur89}).  We conclude that
the existing data strongly supports our hypothesis that the X-ray
outbursts of these two BHSXTs showed almost identical evolution.
However, our goal is to adapt the \acite{emn97} model to \a0, and
since we cannot convert the existing \a0 data to physical flux units
(or at least not enough of it to be of use) for comparison with the
model, we have to obtain some sort of quantitative mapping between
\nm91 and \a0 lightcurves.  If this can be done, then we can assume
that the model for \nm91 (with properly adjusted parameters, as
deduced from the lightcurve mapping) reproduces the behavior of \a0 in
the X-ray band and proceed to test it against the optical data.

\acite{emn97} adopted the following mapping relation between the time
after the outburst maximum ($T$, measured in days), mass accretion
rate ($\Mdot$) and the transition radius between the inner hot flow
(ADAF) and the outer thin disk ($\Rtr$):
\be 
\label{T}
T = \tau_{\Mdot} \ln{\left(\frac{\Mdot_{\rm peak}}{\Mdot}\right)} + 
\tau_{\Rtr} \ln{\left(\frac{\Rtr}{3\Rg}\right)}.  
\ee 
At $T=0$, mass accretion rate is at its maximum value of $\Mdot_{\rm
peak}$ and the transition radius is at its minimum value of $3 \Rg$
(the marginally stable orbit for a non-rotating black hole), i.e. most
of the mass is accreting via a thin disk.  As $T$ increases, the mass
accretion rate decreases exponentially (with a characteristic
e-folding time scale $\tau_{\Mdot}$) as the disk empties out, as
argued by \cite{kir98}.  In this regime, a dominant soft component
originates in the thin disk and hard X-ray emission comes from a hot
corona above the disk.  According to the relative importance of the
hard X-ray component, \acite{emn97} identified the resulting spectra
with the VHS (stronger high energy emission) or HS (very weak high
energy emission).

Once $\Mdot$ drops below the critical value for the formation of an
ADAF, $\Mdot_{\rm crit}$, the disk begins to evaporate starting from
its inner edge, forming the hot accretion flow and thereby increasing
the transition radius (on an e-folding time scale $\tau_{\Rtr}$). This
is accompanied by a dramatic change in the X-ray spectrum.  Since now
most of the emission originates in the very hot, rarefied gas filling
the region inside $\Rtr$, the spectrum quickly becomes very hard, with
most of the emission coming out in $\sim 100\keV$ photons, which are
produced via inverse Compton scattering of seed synchrotron and disk
photons.  This regime, when $\Mdot$ through the inner region stays
near its critical value (fueled by the evaporation of the disk, see
\S\ref{disc}) and $\Rtr$ increases with time, was identified by
\acite{emn97} with the Intermediate state (hereafter IS).  In this
state, spectral and temporal properties of BHSXTs shift from those
characteristic of the HS (soft spectrum) to those of the LS (hard
spectrum).  Finally, at some point (perhaps when $\Rtr$ reaches its
value in quiescence) mass accretion rate through the inner region
begins to drop again and the system moves from the LS into quiescence.

In general, there is no reason to expect the parameters in equation
(\ref{T}) to be exactly identical for two systems, even as otherwise
similar as those considered in this paper.  However, in the scenario
proposed by \acite{emn97}, there is one quantity which must remain
relatively constant from system to system, and that is $\mdotcr =
\Mdot_{\rm crit}/\Mdotedd$, where $\Mdotedd = 1.39 \times 10^{18} m
{\rm \ g\,s^{-1}}$ is the Eddington mass accretion rate for a black
hole of mass $m=M/\msun$.  Therefore, we expect that in Eddington
units the X-ray luminosity of BHSXTs in the IS is more or less constant.  

Based on this fundamental assumption, in Figure \ref{xrayltcv} we have
plotted soft X-ray lightcurves for our two BHSXTs, where the \a0 flux
was scaled to obtain the best match in the section of the lightcurves
corresponding to the intermediate state, as marked on the figure.
The fluxes are normalized to Crab units to remove the effect of
different instrumental effective areas.  Then the scaling factor $\eta =
F_{A}/F_{GRS} = 8$ must be due only to the difference in the black hole
masses, binary inclinations and distances to the two systems.
Roughly, we have
\be
\label{eta}
\eta = {F_{A}\over F_{GRS}} \approx \left(m_{A}\over m_{GRS}\right)
\left(\cos{i_{A}}\over \cos{i_{GRS}}\right) \left(d_{GRS}\over
d_{A}\right)^2.  
\ee 
Note that though the dependence of the observed flux on $\cos{i}$ used
here is strictly valid only for an optically thick disk, the disk
emission dominates soft X-ray flux during early part of the IS.
Moreover, near $\mdotcr$ the emission from ADAF also escapes
preferentially along the polar axes, where the gas density is lower
(see \fcitep{nay95} and \scitep{nbm97}).

\centerline{\null}
\includegraphics{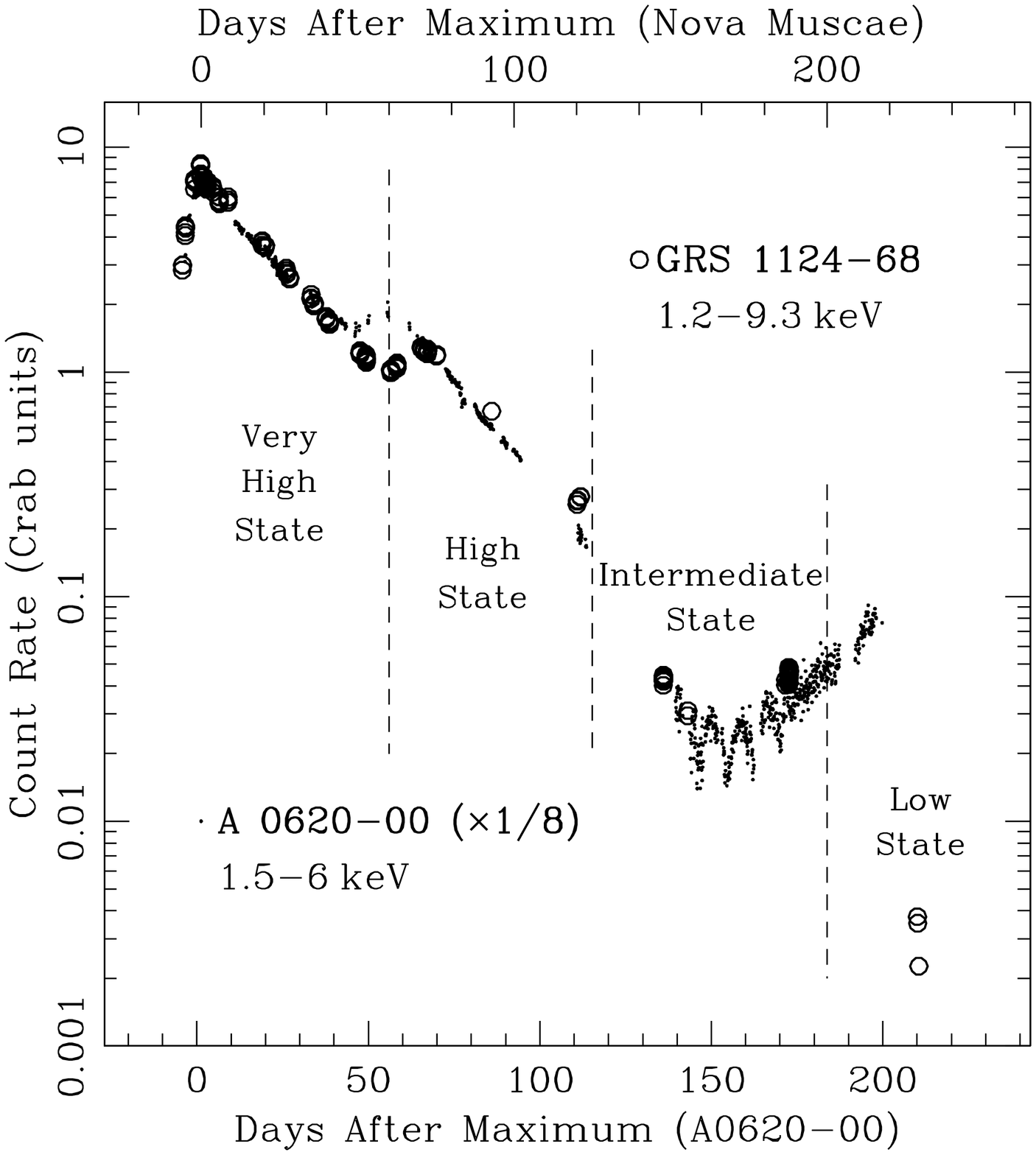} 
\vskip 3.6in 
\figcaption{\label{xrayltcv} Soft X-ray lightcurves of \a0 ({\it SAS-3} 
$CSL\,A$;
$1.5-6\keV$) and \nm91 ({\it Ginga LAC} $1.2-9.3\keV$)
plotted in Crab units.  Different spectral states seen during the
outburst of \nm91 are indicated in the figure.  The count rate for \a0
has been multiplied by a factor of $1/8$, to match the curves in the
IS, when, according to the EMN model, mass
accretion rate is on the order of $\mdotcr$.  Note the close
correspondence between the lightcurves.  They differ only in the time
of the secondary maximum and in the width of the tertiary peak.}
\vskip 0.5cm

\acite{emn97} adopted $m_{GRS} = 6$,
$i_{GRS}=60^{\circ}$ and $d_{GRS} = 5\kpc$ and showed that their model
reproduces the X-ray data of \nm91.  Given these values for \nm91 and
the constraints on the distance to \a0 (see Table \ref{tabbin}),
equation (\ref{eta}) favors a somewhat low black hole mass for \a0.
For $m_{A} = 4.5$, we obtain $d_{A} = 1.4\kpc$ or $1.5\kpc$,
depending on the value of the $f(M)$.  Larger values of $m_{A}$
require placing \a0 at an even greater distance, apparently
inconsistent with existing observations.  An even smaller black hole
mass in \a0, e.g. $m_A = 3.5$, is formally consistent with all the
data; however, there is a preference towards a somewhat higher value.
Therefore, in our model for \a0 we adopt the largest black hole mass
that is still consistent with the distance to the system: $m_A = 4.5;\
i_A= 65^{\circ};\ d_A=1.4\ {\rm kpc}$.

The values of the three parameters for \a0 introduced in equation
(\ref{T}) can be read off Figure \ref{xrayltcv}.  By a direct
comparison of the two outburst time scales, we obtain $\tau_{\Mdot} =
32\,{\rm days}$, quite close to $\tau_{\Mdot} = 35\,{\rm days}$ for
\nm91.  Assuming that at day 200 $\Rtr=10^{3.9}\Rg$ (determined from
observations of H$_{\alpha}$ line in quiescence, see \scitep{nmy96}
and references therein), where $\Rg=2.95\times 10^5 m\ {\rm cm}$ is
the Schwarzschild radius, the length of the intermediate state implies
$\tau_{\Rtr} = 10\,{\rm days}$ (as compared with $\tau_{\Rtr} =
8\,{\rm days}$ for \nm91).  Finally, the ratio of the peak luminosity
to that in the IS is the same in both systems, which implies that the
peak mass accretion rate in \a0 was the same as in \nm91, i.e.
$\mdot_{\rm peak} = \Mdot_{\rm peak}/\Mdotedd = 3$.  Incidentally,
similarity of $\mdot_{\rm peak}$ in these two systems suggests that
like \nm91, \a0 was in the VHS near the peak of the outburst.  The
available spectral information, though limited, also supports this
conclusion.

One must keep in mind however, that our assumption of constant
$\mdotcr$ has been demonstrated to be valid (in theory) only for
accreting black holes with zero spin; and it is possible that the
value of $\mdotcr$ depends on $a/M$.  If the black holes in \a0 and
\nm91 have very different spins, the procedure of matching the two
lightcurves described above may not be valid.  On the other hand,
\fcitet{zcc97} and \fcitet{czc98} argue that \nm91 does not contain a
rapidly spinning black hole, and since \a0 shows similar X-ray
spectra, it is probably not likely to have much spin either.

\section{Modeling UBV Lightcurves of \a0}
\label{opt}

The model of \acite{emn97} underpredicted the optical flux of \nm91 in
the VHS and HS by a factor $\sim 3$, though outer disk irradiation was
self-consistently included in their calculations.  This is not
surprising, since it has recently been demonstrated in several papers
(e.g. see \fcitep{dlhc99} and references therein) that the standard
\fcitet{shs73} thin disks do not provide enough flaring at the outer
edge to reproduce the observed levels of irradiation.  In fact,
self-consistent disk models show that the outer parts of the disk
(even when irradiation is taken into account in the disk height
calculations) are {\em convex}, rather than concave (\fcitep{tmw90};
\scitep{dlhc99}), i.e. they are shielded from the X-rays emitted by
the inner disk.  Since in the luminous spectral states (VHS and HS)
most of the X-ray emission originates in the thin disk, rather than
the corona, this places very severe constraints on how important
irradiation can be in planar disks.

As we will demonstrate below, the ratio of the optical to X-ray fluxes
in \a0 is even greater than that in \nm91.  Clearly then, in order to
reproduce the observed optical emission from the former system, we
need to assume that the surface of the outer disk is considerably more
curved than predicted by the standard model.  An alternative
explanation, involving an extended hot corona which can illuminate
even a concave outer disk, cannot apply when the X-ray spectrum is
dominated by the disk emission and fails to explain the statistics of
disk eclipses in low-mass X-ray binaries (\fcitep{jpa96}).

The well-known equation describing irradiating flux at radius $R$ due
to a compact central source of X-rays is given by
(e.g. \fcitep{shs73}; \fcitep{par96}; \fcitep{kir98}): 
\be 
\label{firr}
F_{\rm irr}(R) = \frac{n L_{\rm X} (1-a)}{4\pi R^2}\left(\frac{H}{R}\right)^n
\left[\frac{d \ln{H}}{d \ln{R}} - 1\right],
\ee

where $L_{\rm X}$ is the total central X-ray luminosity, $a$ is the
disk albedo, and $H$ is the height of the disk photosphere above the
orbital plane.  The power index $n$, introduced by \fcitet{kir98}, is
equal to 1 when the irradiating source radiates isotropically (e.g. as
in the case of a corona); and $n$ is set to 2 when the outer disk is
irradiated by the inner disk, which emits primarily in the direction
perpendicular to the disk surface.  The factor of $n$ in the numerator
(significant only in the case when $n=2$) is necessary to ensure the
correct normalization of the total emission from the thin disk.

We know $L_{\rm X}$ from analyzing the X-ray data (see \S\ref{xrays}),
which also allows us to distinguish whether $n=1$ or $n=2$ (or a
combination of both) is applicable.  Unfortunately, the other two
parameters, $H(R)$ and $a(R)$, are not well known.  The former because,
as we argued above, the standard answer cannot reproduce the
observations.  The latter because published theoretical
calculations of X-ray irradiation, from which a value of $a$ can be
derived for a given incident spectrum, generally average over all
incident angles (e.g. \fcitep{maz95}; \fcitep{zkz94}; \fcitep{wlz88}),
while here only grazing incidence X-rays are relevant.  In fact, the
values of $a$ inferred from observations of persistent low-mass X-ray
binaries with neutron star primaries (\scitep{jpa96}; \fcitep{krv91})
are very close to unity, quite different from the generally accepted
range, $a=0.1-0.2$

We simplify the problem by lumping the two uncertainties together as a
single irradiation parameter ${\cal C}(R)$ (\scitep{dlhc99};
\fcitep{elh99}), so that 
\be 
F_{\rm irr} = {\cal C} (R) \frac{L_{\rm X}}{4 \pi R^2}. 
\label{C} 
\ee (Note that our ${\cal C}$ is defined through $L_{\rm X}$, while
\scitet{dlhc99} define their parameter through $\dot M c^2$, so that
the two differ by the factor $\eta= 0.1$, which represents the
efficiency of converting mass into X-rays for the thin disk.)
Moreover, we assume that the only radial dependence of ${\cal C}$
comes from $H(R)$, for which we adopt a rather ad hoc prescription: $H
\propto R^{9/7}$.  In systems in outburst, the optical band generally
falls in the Rayleigh-Jeans regime of the outermost disk radius.
Therefore, the total optical emission is well described simply by the
value of $H$ at the outer disk edge, $H (R_{\rm out})$, and the exact
form of $H(R)$ affects only the details of the optical spectra.  We
picked the prescription above derived for isothermal disks
(\fcitep{vrg90}), solely to make the connection with previous
literature on disk irradiation, though we know that the outer disks in
\a0 and \nm91 are {\em not} isothermal.  For the X-ray albedo we adopt
(simply for lack of anything more appropriate) the angle averaged
prescription for the X-ray irradiation of a cold unionized medium
(\scitep{wlz88}) which gives $a \sim 0.1-0.2$, depending on the
spectral distribution of incident photons.  We believe that this is
not a bad approximation, since grazing incidence irradiation is not
likely to increase the value of the albedo above $\sim 0.4$, unless
the outer disk has an ionized atmosphere with optical depth of order
unity (D. Psaltis, private communication; see also \fcitep{bst74}).
Thus the error in the quantity $(1-a)$ in Eq. (\ref{firr}) is at most
$50\%$, and not an order of magnitude, as implied by the albedo values
derived by \scitet{jpa96}. Nevertheless, because of the uncertainty in
this quantity, the values for $H(R_{\rm out})$ quoted below cannot be
taken at face value; they must simply be interpreted as a
measure of ${\cal C}$.

With the prescription for the treatment of outer disk irradiation
outlined above, we use the \acite{emn97} model for X-ray emission to
compute a series of spectra following the evolution of \a0 during the
decline phase of its outburst.  Throughout the calculation the shape
of the outer thin disk was held fixed, and the height of the disk at
the outer edge, $H (R_{\rm out})/R_{\rm out}$, was adjusted to fit the
overall normalization of the observed optical emission.  The
irradiating flux originating in the ADAF was computed using the ray
tracing procedure described in detail in EMN, which is roughly
equivalent to using Eq. (\ref{firr}) with $n=1$.  We calculated the
irradiating flux emitted in the inner parts of the thin disk using
Eq. (\ref{firr}) with $n=2$.  To describe the emission from the thin
disk we used the standard modified black-body spectrum.  The effective
temperature at each radius was computed by equating the black-body
emission to the sum of the internal energy dissipation rate in the
disk and the heating rate due to irradiation (as described in
\acite{emn97}).  Note that in the outer disk, which is primarily
responsible for the observed optical emission, X-ray irradiation of
the disk surface completely dominates over the viscous heating.

\centerline{\null}
\includegraphics{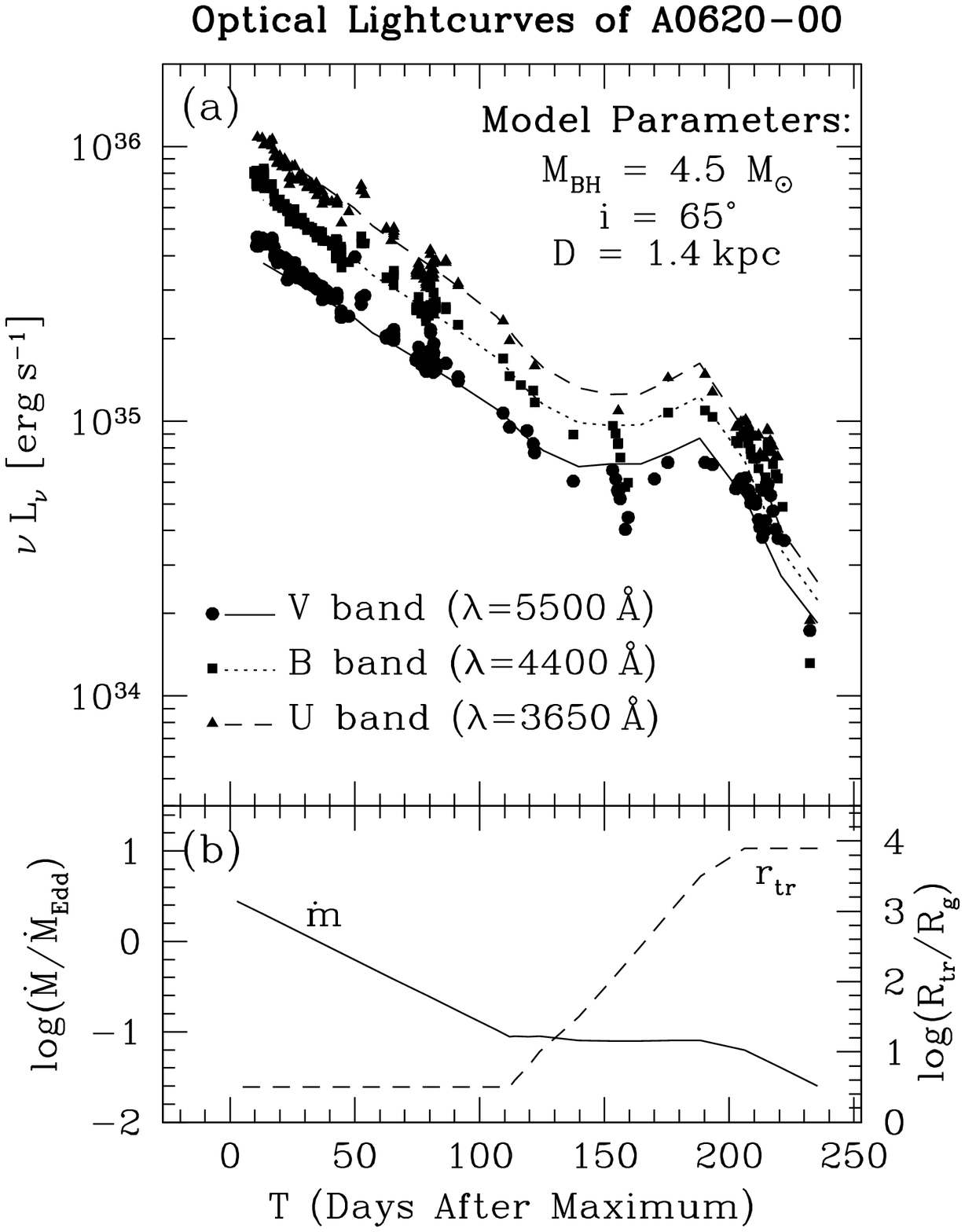} 
\vskip 4.5in 
\figcaption{\label{optltcv}(a) The observed specific luminosity of \a0 in the
three optical bands (Webbink 1978; Kuulkers 1998) is plotted as solid
symbols together with the best fitting model lightcurves, obtained 
with $H(R_{\rm out}) = 0.12 R_{\rm out}$. 
Note that most of the optical flux was generated by the
reprocessing in the outer disk of the X-rays emitted by the inner disk
and the corona (in the VHS and HS) or by the ADAF (in the IS and LS).  
(b) The variation of the mass accretion rate $\mdot = \Mdot/\Mdotedd$
and the transition radius $\rtr=\Rtr/\Rg$ with time in the model.}
\vskip 0.5cm

The resulting optical lightcurves in the standard UBV optical bands,
corresponding to the best fitting value of $H (R_{\rm out}) = 0.12
R_{\rm out}$ are shown in Figure \ref{optltcv}(a).  In the same figure
we plotted the observed optical fluxes for \a0 (from \scitep{web78}
and \scitep{kuu98}), dereddened for $E_{B-V} = 0.35$ (\fcitep{wph83})
and converted to luminosity values assuming the distance to \a0 of
$1.4\,{\rm kpc}$.  

As in modeling the X-ray data, we used the mapping given by equation
(\ref{T}) to construct our lightcurves.  Figure \ref{optltcv}(b) shows
how two main model parameters, $\mdot = \Mdot/\Mdotedd$ and 
$\rtr = \Rtr/\Rg$, change with time 
during the decline phase of the outburst.  The transition radius
changes only during the IS, when mass accretion rate stays constant.
Before and after the IS, $\mdot$ decreases exponentially with time.
Here we take the value of $\rtr$ at the end of the IS to be the same
as that suggested by observations in quiescence (\scitep{nmy96}).
There is no direct observational support for this assumption from the
studies of X-ray spectra, since the shape of X-ray emission stays
essentially fixed for any value of $\rtr$ above $\sim 30$ (see
e.g. \fcitep{enc98}).  However, a comparison of mass accreted during
the IS with the total mass stored in the thin disk (see \S\ref{disc})
favors a large change in $\rtr$.

The general correspondence between the model lightcurves and the data in
Figure \ref{optltcv}(a) is excellent.  Taking into account the quality
of the \acite{emn97} fit to \nm91 X-ray data, one cannot expect to
produce optical lightcurves based on this model that would be in
better agreement with observations.  

Two points are important to emphasize.  First, we reproduce the slower
decay of the optical intensity ($\tau_{\rm opt} \sim 75\,{\rm days}$)
relative to the X-ray intensity ($\tau_{\rm x} \sim 25\,{\rm days}$),
which is a typical feature of optical lightcurves in BHSXTs. Though
optical emission comes purely from X-ray irradiation, cooling of the
outer disk, which accompanies a decline in the irradiating flux,
causes the peak of the emission to shift towards the optical band.
This effect compensates for the total decline in the reprocessed
emission and considerably increases the e-folding time of the observed
optical flux (see also the discussion in \fcitep{kir98}).

The second feature of the observed lightcurves which is reproduced well 
in the model, is the upturn during the IS (between days 130 and 200).  
In our model this is due to the change in the character of the
irradiating flux.  During the early parts of the outburst (VHS and HS)
X-ray emission comes primarily from the accretion disk, so that the
irradiating flux is proportional to the square of the disk height (see Eq.
\ref{firr}).  During this period the irradiation parameter in our
calculations is ${\cal C}(R_{\rm out}) \simeq 0.004$.  On the other
hand, in the LS, when all the X-rays come from the isotropically
emitting ADAF, the irradiating flux is linear in $H/R$.  At this
point, we have ${\cal C}(R_{\rm out}) \simeq 0.03$.  During the IS,
the value of ${\cal C}$ changes gradually between these two limiting
values, causing an upturn in the lightcurve.  This feature may be an
additional confirmation of our assumption that during this part of the
outburst the transition radius is increasing.

Though the overall normalization of the model lightcurves is
determined by our choice of $H(R_{\rm out})$, it is
interesting that the model produces the correct spacing between the V,
B, and U lightcurves, indicating that our optical band spectra must 
agree with the data fairly well.  However, because the functional form of 
$H(R)$ was chosen arbitrarily, this may not be a very 
significant result.  

\section{Revisiting \nm91}
\label{optnm}

As mentioned above, the original \acite{emn97} model fell short (by a
factor of 3) of reproducing the optical emission from \nm91.  However,
with the improved treatment of irradiation we now have no trouble in
fitting those data, as demonstrated in Figure \ref{nmltcrv}.  The UV
data points were taken from \scitet{sgr93}.  The V band data are
collected from \scitet{bai92}; \scitet{khm96}; \scitet{dmb98}; and
other observations tabulated by \acite{chp92} (1992, see references
therein).  \acite{chp92} also obtained an HST spectrum of the system
which we used to estimate the observed flux at $3100 \AA$ for day 127.
All the data points were dereddened using the color index $E_{B-V} =
0.30$ quoted by \scitet{sgr93}. The model lightcurves were calculated
using the parameters for \nm91 given in \acite{emn97}.

\centerline{\null}
\includegraphics{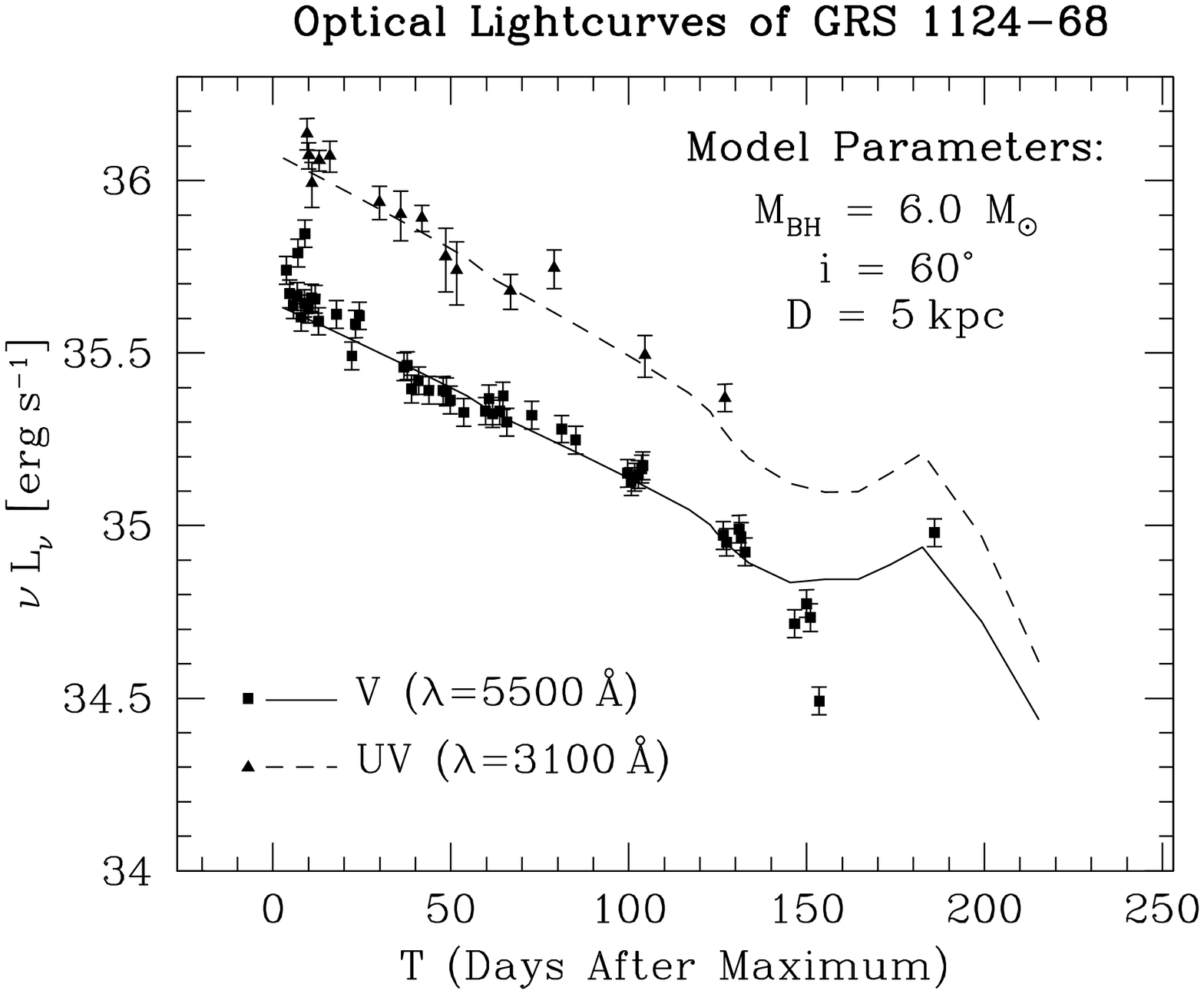} 
\vskip 2.9in 
\figcaption{\label{nmltcrv}  Optical and UV photometry for \nm91 are
plotted as solid points.  Errorbars on the UV points are quoted 
by Shrader \& Gonzalez-Riestra (1993), and those on the V band points 
are all set to $\pm 0.1$ mag.
The curves show the model lightcurves of \nm91 calculated for $\tau_{\Mdot} = 
35\,{\rm days}$ and $\tau_{\Rtr} = 8\,{\rm days}$ (see EMN).
The best fit is obtained with $H (R_{\rm out}) = 0.07 R_{\rm out}$.} 
\vskip 0.5cm

The most interesting point about the fit in Figure \ref{nmltcrv} is
the comparatively small best fit value of the disk height at the 
outer edge, $H(R_{\rm out}) = 0.07 R_{\rm out}$.
Using the fact that most of the data were obtained during the period
when X-ray emission is dominated by the disk, we estimate ${\cal
C}(R_{\rm out}) \simeq 0.0014$.  This value is considerably smaller
than that derived for \a0 in \S\ref{opt}.  It is easy to see the
origin of this discrepancy.  A comparison of Figures \ref{optltcv}(a)
and \ref{nmltcrv} shows that the optical luminosities of the two
BHSXTs were nearly identical.  However \a0 was a factor of 1.6 dimmer
in X-rays (see \S\ref{xrays}), implying that irradiation must have
been considerably more important in this system than in \nm91.

\section{Discussion}
\label{disc} 

Our results show that in order to reproduce the optical lightcurves of
\a0 and \nm91 efficient irradiation of the outer disk is required.
The disk height at the outer edge required to fit the observations is
a factor of $\sim 2$ smaller than $H/R \gsim 0.2$ found by
\scitet{jpa96} for some persistent LMXBs.  However, this comparison of
$H/R$ values is complicated by the fact that the albedo values used in
our calculations were almost an order of magnitude smaller than those
derived by \scitet{jpa96}.  In addition, we assumed a different
angular distribution for the irradiating flux then \acite{jpa96}.  A
better way to check our results would be through comparing the
effective irradiation parameter at the outer disk edge instead of disk
height, since it is really the former that determines the importance
of irradiation.  The results of \scitet{jpa96} imply ${\cal C} (R_{\rm
out}) \sim 0.002-0.004$, consistent with what we find in the VHS and
HS.  On the other hand, our calculations imply that in the IS the
heating of the outer disk is almost an order of magnitude stronger
because of the isotropy of the irradiating X-rays.

The most puzzling result of our investigation is the fact that a
smaller disk in \a0 is heated considerably more than its larger
counterpart in \nm91.  This conclusion is nearly model independent,
since it follows simply from comparing the X-ray to optical flux
ratios in the two systems.  Of course our main assumption that the mass
accretion rate during the IS is the same for \a0 and \nm91 may be 
incorrect, since we do not know precisely what mechanism is driving
the evaporation of the thin disk at this stage of the outburst, or
whether the two black holes have the same spin.  However, even
abandoning this fundamental constraint would not reconcile the
results for the two systems.  If we 
allow $\mdot$ in the IS to be smaller than $\mdotcr$ in \a0, X-ray
data allows higher $m$ and smaller $i$ for this system (which
incidentally are also preferred by observations in quiescence, see
Table \ref{tabbin}).  The size of the disk in \a0 would not be
strongly affected since it depends only weakly on $m$ (see
Eq. \ref{rout}), so most of the effect would come from the change in
$i$.  In a lower inclination system the ratio of optical emission to
X-rays is greater, as is observed in \a0, but only if X-ray emission
is close to isotropic.  Thus, this would have some effect in the late
IS and LS, where X-ray emission originates in an
ADAF, but will not help in the VHS and HS, were X-rays are emitted
predominantly in the disk.

Thus, our comparison of \a0 and \nm91 clearly implies that the disk in
the former is for some reason more `curved', though in all other respects
the two systems are nearly identical.  This suggests that the shape of
the disk in a BHSXT may not be entirely predictable for a given set of
parameters, or in other words, that it might depend on the past
history of the system.  The most obvious example of such an effect is
disk warping (e.g. \fcitep{mbp96}).  If the outer disks in BHSXTs (and
LMXBs in general) are warped, this can explain both high levels of
irradiation and variations between otherwise similar sources, like
those we have examined here.

One constraint we can place on disk warping comes from expected
luminosity fluctuations due to the long-term precession of the warp.
For the simplest warp geometry ($m=1$ mode) we can estimate the ratio
of fluxes observed when the warp is facing the observer and when it is
facing away from the observer as $\cos{(i-\beta)}/\cos{(i+\beta)}$,
where $\beta$ is the angle between the normal to the disk surface at
the outer edge and the binary axis, and $i$ is the binary inclination.
For our prescription $H \propto R^{9/7}$, we have $\beta =
\tan^{-1}{(dH/dR)} \sim 9^{\circ}$; in general for small warp
amplitude $\beta$ is of the same order of magnitude as $H/R$.  Thus,
for our choice of parameters, the optical emission observed from a
warped disk in \a0 must vary by a factor of $\sim 2$ on the time scale
of several months (see e.g. \scitep{mbp96} for radiation-driven
warping), while the emission from \nm91 should vary by about
$50\%$. (Note that no eclipses of either the secondary or the central
X-ray source are expected in either \a0 of \nm91, since in both
systems $\beta+i \ll \pi/2$.)  This number of course must be treated
as only a rough estimate, as it is very sensitive to the exact shape
of the disk surface.  However, it is interesting to note that both
optical and X-ray lightcurves of \a0 show $\sim$7-8 day oscillations
of roughly the right amplitude (though the period seems to be too
short for precession) during the IS (e.g. \fcitep{tmt77};
\fcitep{mbb76}; see also Figures \ref{xrayltcv} and \ref{optltcv}a),
and a hint of the same effect can be seen in the optical lightcurve of
\nm91 (Figure \ref{nmltcrv}, see also the discussion in \scitep{kuu98}).

The \acite{emn97} model does not explain the origin of the secondary
maximum seen near day 55 in the lightcurves of \a0 and around day 70
for \nm91.  Among several possible explanations for this feature that
were proposed in the literature (see \fcitep{can98} and references
therein), two scenarios invoke an enhancement in the central mass
accretion rate triggered by irradiation of a new mass source.
\facite{aks93} (1993, see also \fcitep{clg93}) propose the secondary
star as the source of the extra mass; \scitet{kir98} invoke the outer
disk, initially left cold by the heating wave.  Both mechanisms would
cause an increase in the lightcurve decay time $\tau_{\rm v}$ after 
the peak of the outburst, where $\tau_{\rm v}$ is essentially of 
order of the viscous time
scale in the outer disk.  In \a0, a system with a much smaller disk,
this time scale should be smaller as well, consistent with what is
observed.

The plateau before the tertiary maximum in the X-ray lightcurves, which we 
identify with the end of the IS, was much more prominent in
\a0 than in \nm91.  In our model this means that the phase of 
constant $\Mdot$ and increasing $\Rtr$ lasted roughly $\sim 20$ days longer
in \a0.  An interesting question to consider is why $\Mdot$ in the 
X-ray producing region stays constant
during this period.  One obvious possibility is that this results
directly from disk evaporation.  We can write down the equation for 
the mass accretion rate through the central ADAF during the IS:
\be 
\label{mdotadaf}
\Mdot_{\rm ADAF} = \Mdot_{\rm disk} + \Sigma_{\rm disk} (\Rtr)\ 2 \pi 
\Rtr \dot \Rtr.
\ee 
This simply states that the mass flow through the ADAF is equal to the
mass flow through the disk plus the rate of mass evaporation from the
inner part of the disk.  In outburst, we expect $\Mdot_{\rm disk}$ to
decrease exponentially with time.  Then one can show that if
$\Mdot_{\rm ADAF}$ is kept constant and the surface density profile in
the outer disk is taken to be that of a steady-state disk, the
transition radius has to increase roughly exponentially with time.
Note that we can assume that $\Mdot_{\rm ADAF}$ is constant with $R$
and is equal to the central $\Mdot$, since the accretion timescale in
the ADAF region is many orders of magnitude shorter that the timescale
on which the quantities on RHS of Eq. (\ref{mdotadaf}) vary, i.e. the
ADAF adjusts practically instantaneously to the changes in the mass
inflow at the transition radius.

Though a realistic functional form of $\Rtr (t)$ is probably more
complicated than a simple exponential adopted in Eq. (\ref{T}) (since
it depends on the detailed evolution of the surface density in the
outer disk, which can be obtained only from numerical disk-instability
models), the main results of our paper are not very sensitive to this
detail.  For our model lightcurves to match optical data for \a0, it
is sufficient for $\Rtr$ to increase in some fashion from $3\Rg$ to
$\sim 10^2 - 10^4 \Rg$ during the IS.

If most of the mass accreted onto the black hole during the IS
comes from disk evaporation, it is important to compare the mass
accreted during this period to the mass stored in the thin disk at the
onset of the IS.  The former is simply
\be
M_{\rm acc} = \Mdot_{\rm crit}\ \Delta t_{\rm IS},  
\ee
where $\Delta t_{\rm IS}$ is the duration of the IS.  To 
estimate the mass stored in the disk we make use of the well known fact that 
during the outburst the disk adopts a quasi-steady density profile, so that
\be
M_{\rm disk} = 2 \pi \int_0^{R_{\rm q}} {\Sigma(\Mdot_{\rm crit},R)\ R dR},    
\ee
where $\Sigma(\Mdot_{\rm crit},R)$ is given by the standard outer disk 
formula from 
\fcitet{shs73}, and $R_{\rm q}$ is the value of the transition radius 
at the end of the IS.  For \a0 we get $M_{\rm acc} \sim 
3\times 10^{24}\,{\rm g}$ and $M_{\rm disk} \sim 6\times 10^{23}
(R_{\rm q}/10^{3.9} R_{\rm g})^{5/4} (0.1/\alpha)^{4/5}\,{\rm g}$, where $10^{3.9} R_{\rm g}$ is
the value of the transition radius in quiescence (see \scitep{nmy96}).
Because these numbers are very approximate, the discrepancy between them 
is not serious.  However, we can safely draw the conclusion that 
without any other source of mass, during the IS the transition radius 
must have increased close to its quiescent value.

According to this interpretation, the outer disk contains very little
mass at the end of the IS plateau.  This is consistent with the sharp
drop in the X-ray (see \scitep{kuu98} and references therein) and
optical (see Figure \ref{optltcv}a) fluxes observed at that point.
Moreover, the last set of optical data points in Figure
\ref{optltcv}(a), observed on day 232, shows that the B band flux was
considerably lower than both V and U band fluxes (this was originally
pointed out by \fcitep{lyu76} and \fcitep{lys79}), implying that
perhaps the disk was becoming optically thin at this point.

Though the IS plateau was somewhat shorter in \nm91, the same
calculation for this system gives us very similar values: $M_{\rm acc}
\sim 4\times 10^{24} \,{\rm g}$ and $M_{\rm disk} \sim 10^{24} (R_{\rm
q}/10^{3.9} R_{\rm g})^{5/4} (0.1/\alpha)^{4/5}\,{\rm g}$.  These
numbers are clearly at odds with the results of \fcitet{zds98} which
seem to indicate that in \nm91 the transition radius was still fairly
small at the end of the IS.

\vfill
\section{Conclusions}
\label{conc}

We have tested whether the model proposed by \fcitet{emn97} to explain
the X-ray behavior of \nm91 during its 1991 outburst can also
reproduce the observed optical lightcurves of another BHSXT, \a0.  Our
comparison of the X-ray data for \nm91 and \a0 strongly suggested that
these X-ray novae must have followed the same spectral evolution
during their outbursts and therefore the scenario developed for \nm91
could be directly used in modeling \a0.  We showed that the
\acite{emn97} model modified to include a flared outer disk, can
reproduce the shape of the optical lightcurves of \a0.  Assuming that
the shape of the irradiated outer disk remains fixed throughout the
outburst, this result provides important evidence in support of our
basic scenario of X-ray production, since it relies on the fact that
the primary source of X-rays changes from the inner part of the disk
in the Very High and High spectral states to the isotropically
emitting ADAF in the Intermediate and Low states.

We estimated the strength of irradiation required to reproduce the
observed X-ray to optical flux ratio in \a0, and found that we need an
irradiation parameter (as defined by Eq. \ref{C}) of ${\cal C} \sim
0.004$ when the X-ray production is dominated by the disk, and ${\cal
C} \sim 0.03$ when the irradiating flux comes from the ADAF.  This
corresponds to a fixed disk height at the outer edge of $H/R \sim
0.12$, though this number probably should not be taken at face value
due to considerable uncertainty in the value of the X-ray albedo
of the outer disk.  However, for \nm91 we estimated that the optical
data taken during the VHS and HS requires an irradiation parameter a factor
of $\sim 3$ smaller than in \a0, corresponding to $H/R \sim 0.07$.  
We speculate that this difference between otherwise nearly identical 
systems implies that outer disk flaring is caused by warping.

\acknowledgements We thank Ken Ebisawa for providing the raw count
rates for \nm91 as well as the relevant Crab count rates for Ginga
LAC.  We acknowledge the use of the processed SAS-3 data from Kenneth
Plaks, Jonathan Woo and George Clark.  AE gratefully acknowledges
useful discussions with Kristen Menou, Jean-Marie Hameury and Jean
Pierre Lasota, who significantly enhanced her understanding of the
disk instability model.  This work was supported by NASA through
Chandra Postdoctoral Fellowship grant \#PF8-10002 awarded by the
Chandra X-Ray Center, which is operated by the SAO for NASA under
contract NAS8-39073.  RN was supported in part by grant AST 9820686
from the NSF.

\bibliography{adaf}

\vfill\eject
\begin{deluxetable}{lcc}
\tablecaption{Summary of the Binary Parameters for GRS~1124$-$68 and 
A~0620$-$00 \label{tabbin}} 
\tablehead{Source Name & GRS~1124$-$68\tablenotemark{a} & A0620-00}
\startdata
Orbital Period, $P_{\rm orb}$ (hr) & 10.4 & 7.75\tablenotemark{e} \\
Mass Function, $f(M/M_{\odot})$ & $3.01 \pm 0.15$, $3.34\pm0.15$\tablenotemark{b} & $2.91 \pm 0.08$\tablenotemark{d}, $2.72\pm 0.06$\tablenotemark{f} \\
Companion Type & K3-5V & K5V \\
Black Hole Mass, $m = M/M_{\odot}$ & $6.0^{+1.5}_{-1.0},
\ \ 5.8^{+4.7}_{-2}$\tablenotemark{c} & 
$10\pm7$\tablenotemark{g}, $4.6\pm1.0$\tablenotemark{h} \\
Mass Ratio, $q = M_c/M$ & $0.133\pm 0.019$\tablenotemark{d},
$\ \ 0.128\pm0.04$\tablenotemark{b}& $0.067\pm 0.01$\tablenotemark{f}, 
$> 0.094$\tablenotemark{h}\\
Inclination, $i$ & $60^{+5^{\circ}}_{-6^{\circ}},\ \ 
54^{+20^{\circ}}_{-15^{\circ}}$\tablenotemark{c}& 
$37^{+28^{\circ}}_{-3^{\circ}}$\tablenotemark{g, k}, 
$\ \ 67\pm 4^{\circ}$\tablenotemark{h}\\
Outer Disk Radius\tablenotemark{1}, $R_{\rm out}$ ($10^{11}$ cm) & 
$1.9\pm0.16$ & $1.2\pm 0.27$ \\ 
Distance, $d$ (kpc) & $5.0\pm 1.1$ & $1.05 \pm 0.4$\tablenotemark{g,m} \\
\enddata

\tablenotetext{1}{Computed using equation (\ref{rout}).}
\tablerefs{(a) From \acite{emn97} and references therein, except where 
explicitly indicated; (b) \scitep{cmc97}; (c) \scitep{snc97}; 
(d) \scitep{obr94}; (e) \scitep{mcr86}; (f) \scitep{mrw94}; 
(g) \scitep{snc94}; (g) \scitep{hrh93}; (k) \scitep{sbc99}; 
(m) \scitep{bmg96}.}

\end{deluxetable}

\end{document}